\documentclass{llncs}
\usepackage{authblk}

\usepackage[utf8x]{inputenc}
\usepackage[english]{babel}
\usepackage{subfigure}
\usepackage{float}
\usepackage{placeins}
\usepackage{graphicx}
\usepackage{array}
\usepackage{setspace}
\usepackage{xspace}
\usepackage{cite}
\usepackage[show]{notes}

\usepackage{hyperref}

\newcommand{\AlgFont}[1]{\textbf{#1}}

\newcommand{\GB}{\AlgFont{GB}\xspace}
\newcommand{\ADA}{\AlgFont{ADA}\xspace}
\newcommand{\BAG}{\AlgFont{BAG}\xspace}
\newcommand{\DT}{\AlgFont{DT}\xspace}
\newcommand{\RF}{\AlgFont{RF}\xspace}
\newcommand{\NAIVE}{\AlgFont{NAIVE}\xspace}
\newcommand{\MNB}{\AlgFont{MNB}\xspace}
\newcommand{\LOG}{\AlgFont{LOG}\xspace}
\newcommand{\COMB}{\AlgFont{COMB}\xspace}

\newcommand{\MeasuresFont}[1]{\textbf{#1}}

\newcommand{\TP}{\MeasuresFont{TP}}
\newcommand{\FP}{\MeasuresFont{FP}}
\newcommand{\TN}{\MeasuresFont{TN}}
\newcommand{\FN}{\MeasuresFont{FN}}
\newcommand{\Precision}{\MeasuresFont{Pr}}
\newcommand{\Recall}{\MeasuresFont{Re}}
\newcommand{\FOne}{\MeasuresFont{F1}}
\newcommand{\TypeOne}{\MeasuresFont{Type-I}}
\newcommand{\TypeTwo}{\MeasuresFont{Type-II}}
\newcommand{\Bacc}{\MeasuresFont{BACC}}

\title{Firms Default Prediction with Machine Learning}
\author{Tesi Aliaj\inst{1} \and Aris
  Anagnostopoulos\inst{1}\thanks{Partially supported by ERC Advanced
    Grant  788893 AMDROMA ``Algorithmic and Mechanism Design Research in Online Markets.''} \and Stefano
  Piersanti\inst{1}\inst{2}}

\institute{
Sapienza University of Rome, Italy.
  \email{tess.aliaj@gmail.com, aris@diag.uniroma1.it, piersanti@diag.uniroma1.it}
\and Bank of Italy}

\begin{document}

\maketitle

\begin{abstract}
Academics and practitioners have studied over the years 
models for predicting firms bankruptcy, 
using statistical and machine-learning approaches.
An earlier sign that a company has financial difficulties and may
eventually bankrupt is going in \emph{default}, which, loosely speaking
means that the company has been having difficulties in repaying its
loans towards the banking system.
%
Firms default status is not technically a failure but is very
relevant for bank lending policies and often anticipates the failure of
the company.
Our study uses, for the first time according to our
knowledge, a very large database of granular credit data from the
Italian Central Credit Register of Bank of Italy that contain
information on all Italian companies' past behavior towards the entire Italian banking system
to predict their
default using machine-learning techniques. 
Furthermore, we combine these data with other information
regarding companies' public balance sheet data.
We find that ensemble techniques and random forest provide the best
results, corroborating the findings of Barboza et al. (Expert Syst.
Appl., 2017).
\end{abstract}


\section{Introduction}
\label{sec:intro}

Bankruptcy prediction of a company is, not surprisingly, a topic that
has attracted a lot of research in the past decades by multiple
disciplines~\cite{altman-bankruptcy-17,kumar-review-07,chen-bankruptcy-11,lee-bankruptcy-13,erdogan-bankruptcy-13,cho-bankruptcy-10,wang-bankruptcy-11,Altman-8,Ohlson-9,Begley-10,Lee-10a,Fernandez-11,Odom-13,Atiya-15,Wang-16}.
Probably the main importance of such research is in bank lending.
Banks need to predict the possibility of default of a potential
counter-party before they extend a loan.
An effective predictive system can lead to a sounder and profitable
lending decisions leading to significant savings for the
banks and the companies and, most importantly, to a stable financial
banking system.
A stable and effective banking system is crucial for
financial stability and economic recovery as well highlighted by the
recent global financial crisis and European debt crisis.
According to Fabio Panetta, general director of the Bank of Italy,
referring to Italian loans,
``The growth of the new deteriorated bank loans and the slowness of the
judicial recovery procedures have determined a rapid increase in the
stock of these assets, which in 2015 reached a peak of 200 billion,
equal to 11 percent of total loans.''\footnote{Fabio Panetta, Chamber of Deputies, Rome, May 10, 2018.}


Of course, despite the plethora of studies, predicting the failure of a
company is a hard task, as demonstrated by the 
enormous increase in large corporate failures in
the last decades.

%
%
%
%
%

Most related research has focused on \emph{bankruptcy} prediction, which
takes place when the company officially has the status of being unable
to pay its debts (see Section~\ref{sec:problem}). However, companies
often signal much earlier their financial problems towards the banking
system by going in \emph{default}. Informally speaking, a company enters
into a default state if it has failed to meet its requirement to repay
its loans to the banks and it is very probably that it will not be able to meet his financial commitments in the future
(again, see Section~\ref{sec:problem}). Entering into a
default state is a strong signal of a company's failure: typically banks
do not finance a company into such a state and it is correlated with
future bankruptcy.

In this paper we use historic data for predicting whether a company will
enter in default. We base our analysis on two sets of data. First, we
use historic information from \emph{all the loans} obtained by \emph{almost all
the companies} based in Italy (totaling to around $800K$ companies).
This information includes information on the companies credit dynamics in the past
years, as well as past information on relations with banks and on values
of protections associated to loans.
Second, we combine these data with the balance sheets of $300K$
of these companies (the rest of them are not obliged to produce balance
sheets). We apply multiple machine-learning techniques, showing that the
future default status can be predicted with reasonable accuracy.
Note that the dimensions and the information in our dataset exceeds
significantly those of past work, allowing to obtain a very accurate
picture of the possibility to predict over various economic sectors.

\paragraph{Contributions.} To summarize the contributions of our paper
\begin{enumerate}
\item We analyze a vary large dataset ($800K$ companies) with highly granular
data on the performance of each company over a period of
10 year. To our knowledge, this is the most extensive dataset used in
the literature.
\item We use these data to predict whether a company will default in the
next year.
\item We combine our data with data available from company balance
sheets, showing that we can improve further the accuracy of predictions.
\end{enumerate}

\paragraph{Roadmap.} In Section~\ref{sec:related} we present some related
work. In Section~\ref{sec:problem} we provide definitions and we
describe the problem that we solve. In Section~\ref{sec:approach} we
describe our datasets and the techniques that we use and in
Section~\ref{sec:experiments} we present our results. We conclude in
Section~\ref{sec:conclusion}.

\section{Related Work}
\label{sec:related}

There has been an enormous amount of work on bankruptcy prediction. Here
we present some of the most influential studies.

Initially, scholars focused on making a linear distinction among healthy
companies and the ones that will eventually default. Among the most
influencing pioneers in this field we can distinguish
Altman~\cite{Altman-8} and Ohlson \cite{Ohlson-9}, both of whom made a
traditional probabilistic econometric analysis. Altman, essentially
defined a score, the $Z$ discriminant score, which depends on several
financial ratios (working capital/total assets, retained earnings/total
assets, etc.) to asses the financial condition of a company.
Ohlson on the other side, is using a linear
regression (LR) logit model that estimates the probability of failure of
a company.
%
Some papers papers criticize these methods as unable to classify
companies as viable or nonviable~\cite{Begley-10}. However, both
approaches are used, in the majority of the literature, as a benchmark
to evaluate more sophisticated methods.

Since these early works there has been a large number of works based on
machine-learning techniques \cite{lin-12, devi-18, nanni-09}.
The most successful have been based on 
decision trees \cite{Lee-10a, Zhou-10b, Gepp-10b, Martinelli-12}
and neural networks
\cite{Fernandez-11, Odom-13, Boritz-14,Atiya-15,Wang-16}. Typically, all
these works use different datasets and different sets of features,
depending on the dataset.
Barboza et al.~\cite{altman-bankruptcy-17} compare such techniques with
support vector machines and ensemble methods showing that ensemble
methods and random forests perform the best.

These works mostly try to predict bankruptcy of a company. Our goal is
to predict default (see Section~\ref{sec:problem}). Furthermore, most of
these papers use balance-sheet data (which are public). Our dataset
contains a granular information of a very large set of companies on the
past behavior of loan repayment. To our knowledge, this is the most
extensive dataset used in the literature.

\section{Firm-Default--Prediction Problem}
\label{sec:problem}

There are many technical terms used to characterize debtors who are in
financial problems: illiquidity, insolvency, default, bankruptcy, and so
on. Most of the past research on prediction of failures
addresses the concept of \emph{firm bankruptcy}, which is the legal
status of a company, in the public registers, that is unable to pay its
debts. A firm is in \emph{default} towards a bank, if it is unable to meet
its legal obligations towards paying a loan. There are specific
quantitative criteria that a bank may use to give a default status to a
company.

%

\subsection{Definition of \emph{Adjusted Default Status}}
\label{subsec:adjusted-default}

The recent financial crisis has led to a revision and harmonization at
international level of the concept of loan default. In general, default
is the failure to pay interest or principal on a loan or security when
due.
In this paper we consider the classification of \emph{adjusted default
status}, which is a classification that the Italian Central bank (Bank of Italy) gives to a company that
has a problematic debt situation towards the entire banking system.
%
It represents a supervisory concept, whose aim is to extend the default credit status to
all the loans of a borrower towards the entire financial system (banks,
financial institutions, etc.).
The term refers to the concept of the Basel II international accord of
\emph{default} of customers.
According to this definition, a borrower is defined in default if its
credit exposure has became significantly negative.
In detail, to asses the status of adjusted default, Bank of Italy
considers three types of negative exposures. They are the following, in
decreasing order of severity:
(1) A \emph{bad (performing) loan} is the most negative classification;
(2) an \emph{unlikely to pay} (UTP) loan is a loan for which the bank
has high probability to loose money; (3) A loan is \emph{past due} if it
is not returned after a significant period past the deadline.

Bank of Italy classifies a company in \emph{adjusted default}, or
\emph{adjusted non performing loan} if
it has a total
amount of loans belonging to the aforementioned three categories exceeding
certain pre-established proportionality thresholds~\cite{adjusted-default-def}.
%
Therefore, a firm's adjusted default classification derives from
quantitative criteria and takes into account the company's debt exposure
to the entire banking system.

If a company enters into an adjusted-default status then it is typically
unable to obtain new loans. Furthermore, such companies are multiple
times more likely to bankrupt in the future. For instance, out of the
$13K$ companies that were classified in a status of adjusted default
in December of 2015, $2160$ ($16.5\%$) 
were no longer active in 2016, having gone bankrupt or being in another
similar bad condition.
On the other hand, only $2.4\%$ of the companies that were not in adjusted
default status became bankrupt.



In this paper we attempt to predict whether a company will obtain an
adjusted default status, although for brevity we may call it just
default.

%

\section{Data and Methods}
\label{sec:approach}

In this section we describe the data on which we based our analysis and
the machine-learning techniques that we used.

\subsection{Dataset Description}
\label{subsec:dataset-description}

Our analysis is based on two datasets. The first and most important in
our work is composed of information on loans and the credit of a large
sample of Italian companies. The second reports balance sheet data of a
large sub-sample of medium-large Italian companies.

\paragraph{Credit data.} 

The first dataset
consists of a very large and high granular dataset of credit
information about Italian companies belonging to the Italian central
credit register (CCR).  It is an information system on the debt
of the customers of the banks and financial companies supervised by the
Bank of Italy. Bank of Italy collects information on customers'
borrowings from the intermediaries and notifies them of the risk
position of each customer vis-à-vis the banking system.
By means of the CCR the Bank of Italy provides
intermediaries with a service intended to improve the quality of the
lending of the credit system and ultimately to enhance its stability.
The intermediaries report to the Bank of Italy on a monthly basis the
total amount of credit due from their customers: data information about
loans of $30,000$ euro or more and non-performing loans of any amount.
The Italian CCR has three main goals: (1)
to improve the process of assessing customer creditworthiness, (2) to
raise the quality of credit granted by intermediaries, and to (3) strengthen
the financial stability of the credit system. 

The crucial feature of this database is the high granularity of credit
information.
It contains information for about $800K$ companies for each quarter of
the the period of 2009--2014. The main features are shown in
Table~\ref{tbl:attributes}.
%


\begin{table}
\small
\begin{center}
\begin{tabular}{| lllc | lllc |}
\hline
& & & & & & &  \\
   &  \textbf{ID}       &\textbf{Description} &  & & \textbf{ID}       &\textbf{Description}   &\\
\hline
\hline
&\textbf{L1}    &Granted amount of loans & & &\textbf{B1}  &Revenues    &  \\
&\textbf{L2}    &Used amount of loans & & &\textbf{B2}  &ROE   &  \\
&\textbf{L3}    &Bank's classification of firm & & &\textbf{B3}  &ROA    &  \\
&\textbf{L4}    &Average amount of loan used& & &\textbf{B5}  &Total turnover    & \\
&\textbf{L5}    &Overdraft & & &\textbf{B6}  &Total assets     & \\
&\textbf{L6}    &Margins & & &\textbf{B7}  &Financial charges/operating margin     & \\
&\textbf{L7}    &Past due (loans not returned after the deadline) & & &\textbf{B8}  &EBITDA   & \\
&\textbf{L8}    &Amount of problematic loans & & &  &  & \\
&\textbf{L9}    &Amount of non-performing loans & & &  &  & \\
&\textbf{L10}    &Amount of loans protected by a collateral & & &  &  & \\
&\textbf{L11}    &Value of the protection & & &  &  & \\
&\textbf{L12}    &Amount of forborne credit & & &  &  & \\
\hline
\end{tabular}
\\
\caption{Main attributes for the loan (L) and the balance-sheet (B)
datasets.}
\label{tbl:attributes}
\end{center}
\end{table}

\paragraph{The Balance-sheets dataset.}

Our second dataset consists of the balance-sheet data of about $300K$
Italian firms. They are generally medium and large companies and they
form a subset of the $800K$ companies with loan data. The main features
include those that regard the profitability of a company, such as return
of equity (ROE) and return of assets (ROA); see Table~\ref{tbl:attributes} for a more
extended list. Typically balance sheet data are public data and have been used extensively for
bankruptcy prediction (e.g., see Barboza et
al.~\cite{altman-bankruptcy-17} and references therein).
%
%
%
%

\subsection{Machine-Learning Approaches}
\label{subsec:algorithms}

As we explain in Section~\ref{sec:related}, the first approaches for
assessing the likelihood of companies to fail were based on some fixed
scores; see the work by Altman~\cite{Altman-8}. Current approaches are
based on more advanced machine-learning techniques. In this paper we
follow the literature~\cite{altman-bankruptcy-17} by considering a set
of diverse machine-learning approaches for predicting loan defaults.

In the first test we used five well-known machine-learning approaches.
We provide a brief description of each of them, as provided by
Wikipedia.

\textbf{Decision Tree (\DT)}: one of the most popular tool in decision
analysis and also in Machine Learning. A decision tree is a
flowchart-like structure in which each internal node represents a
``test'' on an attribute, each branch represents the outcome of the test, and
each leaf node represents a class label (decision taken after computing
all attributes). The paths from root to leaf represent
classification rules.

\textbf{Random Forest (\RF)}: Random forest are an ensemble learning
method for classification, regression and other tasks, that operate by
constructing a multitude of decision trees at training time and
outputting the class that is the mode of the classes. Random decision
forests correct for decision trees' habit of overfitting to their
training set.

\textbf{Bagging (\BAG)}: Bootstrap aggregating, also called bagging, is a
machine learning ensemble meta-algorithm designed to improve the
stability and accuracy of machine learning algorithms used in
statistical classification and regression. It also reduces variance and
helps to avoid overfitting. Although it is usually applied to decision
tree methods, it can be used with any type of method. Bagging was
proposed by Leo Breiman in 1994 to improve classification by combining
classifications of randomly generated training sets.

\textbf{AdaBoost (\ADA)}: AdaBoost, short for Adaptive Boosting, is a
machine learning meta-algorithm formulated by Yoav Freund and Robert
Schapire, in 2003. It can be used in conjunction with many other types
of learning algorithms to improve performance. The output of the other
learning algorithms ('weak learners') is combined into a weighted sum
that represents the final output of the boosted classifier. AdaBoost
(with decision trees as the weak learners) is often referred to as the
best out-of-the-box classifier.

\textbf{Gradient boosting (\GB)}: Gradient boosting is a machine learning
technique for regression and classification problems, which produces a
prediction model in the form of an ensemble of weak prediction models,
typically decision trees. It builds the model in a stage-wise fashion
like other boosting methods do, and it generalizes them by allowing
optimization of an arbitrary differentiable loss function. That is,
algorithms that optimize a cost function over function
space by iteratively choosing a function (weak hypothesis)
that points in the negative gradient direction.

Except for these standard techniques, we also combined the various
classifiers in the following way. After learning two versions
of each classifier, one with the default parameters of the
\texttt{Python scikit} implementation and one with optimal parameters,
we apply all of them (10 in total) and if at least 3 classifiers predict that a firm
will default then the classifier predicts default. The number 3 was
chosen after experimentation. We call this approach combination approach
\textbf{\COMB}.

\section{Experimental Results}
\label{sec:experiments}

The main goal of our study is to evaluate the extent to which we can
predict whether a company will enter in a default state using data from
past years. In particular, our goal is to predict whether a company that
by December 2014 is not in default, will enter in default during one of
the four trimesters of 2015. To do the prediction, we initially used
data from the period of 2006--2014; however we noticed that using data
running earlier than five trimesters before 2015 did not help.
Therefore, for all the experimental results that we report here, we use
the data from the last quarter of 2013 and the four quarter of
2014.

\subsection{Evaluation Measures}

We use a variety of evaluation measures to assess the effectiveness of
our classifiers, which we briefly define. As usually, in a binary
classification context, we use the standard concepts of true positive
(TP), false positive (FP), true negative (TN), false negative (FN):
\medskip

\begin{center}
\begin{tabular}{|l|c|c|}
\hline
	& Predicted Default	& Predicted Not Default \\\hline
Default 	&\TP	&\FN	\\\hline
Did not default &\FP	&\TN	\\\hline
\end{tabular}
\end{center}
\medskip

For instance, FN is the number of firms that defaulted during 2015 but
the classifier predicted that they will not default.

We now define the measure that we use:
\begin{itemize}
\item Precision: $\displaystyle \Precision=\frac{\TP}{\TP+\FP}$
\item Recall: $\displaystyle \Recall=\frac{\TP}{\TP+\FN}$
\item F1-score: $\displaystyle \FOne=2\cdot\frac{\Precision\cdot\Recall}{\Precision+\Recall}$
\item Type-I Error: $\displaystyle \TypeOne=\frac{\FN}{\TP+\FN}$
\item Type-II Error: $\displaystyle \TypeTwo=\frac{\FP}{\TN+\FP}$
\item Balanced Accuracy: $\Bacc \FOne=2\cdot\frac{\TP\cdot\TN}{\TP+\TN}$
\end{itemize}


\subsection{Datasets}

In Section~\ref{subsec:dataset-description} we describe the datasets
that we use. We perform two families of experiments. In the first one,
we use only the loan data (as typically performed by Bank of Italy) to
assess the probability of default. Then, we also combine this information
with balance-sheet data.
Because only a subset of $300K$ (out of the $800K$) companies have balance-sheet data
available, to be able to compare the results, we report here the
findings over this subset for both families of experiments.

\subsection{Balanced versus Imbalanced Classes}

The classification problem that we deal is very imbalanced: around $4.3\%$
of the firms were in a default state in 2015. This common problem in
classification makes it harder. Therefore, as performed in prior
work~\cite{altman-bankruptcy-17} we consider two cases. First we use the
entire dataset, second we also create a balanced version by selecting
all the firms that defaulted ($13.2K$) and an equal number of random
firms that did not default, creating in this way a balanced dataset of
$26.4K$ firms.

\subsection{Baselines}

We evaluated the techniques presented in
Section~\ref{subsec:algorithms}. To assess their effectiveness, we
compare them with three basic approaches.
The first one is a simple multinomial Na\"ive Bayes (\MNB) classifier.
The second is a logistic regression (\LOG) classifier. Finally, we
created the following simple test.
We first measured the correlation of each feature with the target variables
(refer to Table~\ref{tbl:attributes}).
We found the most significant ones, (i.e., the ones that are mostly correlated with the target variable)
are L3 (a bank’s classification of the firm) and L7 (amount of loans not
repaid after the deadline) for the loan dataset and B2 (ROE) and B3
(ROA) for the balance sheet dataset.

Then we built the simple classifier that outputs \emph{default} if at
least one of L3 or L7 are nonzero and \emph{not default} otherwise for the loan dataset. We
call this baseline \NAIVE.

We gather the classification approaches that we use in
Table~\ref{tbl:explanation}.

\begin{table}
\small
\begin{center}
\begin{tabular}{lclc}
\hline
             \textbf{ID method}  &  & \textbf{Description}  \\
\hline
\hline
\NAIVE  &        &Naive classifier based on features correlation with target \\
\MNB    &    &Multinomial Bayesian classifier \\
\LOG     &    &Logistic Regression \\ \hline
\GB    &   &Gradient Boosting \\
\RF    &     &Random Forest  \\
\DT    &      &Decision Tree\\
\BAG  &        &Bagging \\
\ADA  &        &AdaBoost\\\hline
\COMB   &       &Combined method based on multiple classifiers \\
\hline
\end{tabular}
\caption{Baselines and classification algorithms.}
\label{tbl:explanation}
\end{center}
\end{table}

\subsection{Prediction of Adjusted Default}

We are now ready to predict whether companies will enter into an
adjusted default state, as we explain in
Section~\ref{subsec:adjusted-default}.

First we present the results for the original, imbalanced dataset. In
Table~\ref{tbl:unb_loan} we present the results when we use only the
loan dataset, whereas in Table~\ref{tbl:unb_loan_bal} we present the
results when we also use the balance-sheet data. The first finding is
that the evaluation scores are rather low. This is in accordance to all
prior work, indicating the difficulty of the problem.
We observe that the
machine-learning approaches are better than the baselines and the
various algorithms trade off differently over the various evaluation
measures. Random
forests perform particularly well (in accordance with the findings of
Barboza et al.~\cite{altman-bankruptcy-17}) and our
combined approach (\COMB) is able to trade off between precision and
recall and give an overall good classification.
Comparing Table~\ref{tbl:unb_loan} with Table~\ref{tbl:unb_loan_bal} we
see that the additional information provided by the balance-sheet data
helps to improve the classification.

\setlength{\tabcolsep}{12pt}

\begin{table}
\small
\begin{center}
\begin{tabular}{lcccccccccl}
\hline
&\Precision& \Recall  &\FOne & \TypeOne & \TypeTwo  & \Bacc\\
\hline
\hline
\NAIVE          &0.25 &0.11 &0.16       &0.89 &0.04 &0.54 \\
\MNB        &0.95 &0.05 &0.09       &0.95 &0.02 &0.52 \\
\LOG         &0.44 &0.01 &0.02       &0.99 &0.01 &0.50 \\\hline
\GB       &0.63 &0.22 &0.33       &0.78 &0.01 &0.61 \\
\RF         &0.61 &0.21 &0.31      &0.79 &0.01 &0.60  \\
\DT          &0.27 &0.29 &0.28       &0.71 &0.03 &0.63 \\
\BAG          &0.53 &0.19 &0.28      &0.81 &0.01 &0.59  \\
\ADA          &0.56 &0.20 &0.30       &0.80 &0.01 &0.60 \\\hline
\COMB         & 0.52  &0.32  & 0.40      &0.68   & 0.01   &0.66 \\
\hline
\end{tabular}
\caption{Imbalanced training set; loan data.
Higher values are better, except for Type-I and Type-II error.}
\label{tbl:unb_loan}
\end{center}
\end{table}

\begin{table}
\small
\begin{center}
\begin{tabular}{lcccccccccl}
\hline
&\Precision& \Recall  &\FOne & \TypeOne & \TypeTwo  & \Bacc\\
\hline
\hline
\NAIVE          &0.29 &0.14 &0.20       &0.89 &0.06 &0.55 \\
\MNB        &0.95 &0.06 &0.09       &0.95 &0.03 &0.52 \\
\LOG         &0.46 &0.02 &0.03       &0.99 &0.02 &0.50 \\\hline
\GB       &0.63 &0.23 &0.34       &0.77 &0.01 &0.61 \\
\RF         &0.68 &0.25 &0.37      &0.75 &0.01 &0.62  \\
\DT          &0.28 &0.32 &0.30       &0.68 &0.04 &0.64 \\
\BAG          &0.59 &0.21 &0.31      &0.79 &0.01 &0.60  \\
\ADA          &0.61 &0.26 &0.36       &0.74 &0.01 &0.63 \\\hline
\COMB         &0.55 &0.36 &0.43       &0.64 &0.01 &0.67 \\
\hline
\end{tabular}
\caption{Imbalanced training set; loan and balance-sheet data.
Higher values are better, except for Type-I and Type-II error.}
\label{tbl:unb_loan_bal}
\end{center}
\end{table}

In Tables~\ref{tbl:bal_loan} 
and \ref{tbl:bal_loan_bal} we present the results for the balanced
dataset. There are some interesting findings here as well. First, as
expected the classification accuracy improves (similarly to
\cite{altman-bankruptcy-17}). Second, notice that the \NAIVE classifier
performs well (expected, as feature L3 takes into account several factors
of the company's behavior); however the type-II error is high. Overall, \COMB
approach remains the best performer.

\begin{table}
\small
\begin{center}
\begin{tabular}{lccccccccl}
\hline
&\Precision& \Recall  &\FOne & \TypeOne & \TypeTwo  & \Bacc\\
\hline
\hline
\NAIVE          &0.24 &0.78 &0.37       &0.28 &0.50 &0.62 \\
\MNB        &0.43 &0.08 &0.14      &0.88 &0.03 &0.51 \\
\LOG         &0.36 &0.21 &0.26       &0.79 &0.03 &0.59 \\\hline
\GB       &0.23 &0.67 &0.34       &0.33 &0.10 &0.78 \\
\RF         &0.16 &0.73 &0.26      &0.27 &0.17 &0.78  \\
\DT          &0.10 &0.69 &0.17       &0.31 &0.30 &0.69 \\
\BAG          &0.16 &0.69 &0.25      &0.31 &0.17 &0.76  \\
\ADA          &0.24 &0.65 &0.35       &0.35 &0.09 &0.78 \\\hline
\COMB         &0.20   &0.69  &0.31      &0.31   &0.13    &0.78 \\
\hline
\end{tabular}
\caption{Balanced training set; loan data.
Higher values are better, except for Type-I and Type-II error.}
\label{tbl:bal_loan}
\end{center}
\end{table}

\begin{table}
\small
\begin{center}
\begin{tabular}{lccccccccl}
\hline
&\Precision& \Recall  &\FOne & \TypeOne & \TypeTwo  & \Bacc\\
\hline
\hline
\NAIVE          &0.25 &0.77 &0.38       &0.23 &0.49 &0.64 \\
\MNB        &0.44 &0.09 &0.15       &0.91 &0.03 &0.53 \\
\LOG         &0.36 &0.22 &0.28       &0.78 &0.03 &0.60 \\\hline
\GB       &0.19 &0.78 &0.30       &0.22 &0.15 &0.81 \\
\RF         &0.18 &0.80 &0.30      &0.20 &0.16 &0.82  \\
\DT          &0.10 &0.71 &0.18       &0.29 &0.26 &0.72 \\
\BAG          &0.17 &0.75 &0.27      &0.25 &0.17 &0.79  \\
\ADA          &0.18 &0.76 &0.29       &0.24 &0.16 &0.80 \\\hline
\COMB         &0.19   &0.84  &0.31       &0.16   &0.16    &0.84 \\
\hline
\end{tabular}
\caption{Balanced training set; loan and balance-sheet data.
Higher values are better, except for Type-I and Type-II error.}
\label{tbl:bal_loan_bal}
\end{center}
\end{table}

\subsection{A Practical Application: Probability of Default for Loan
Subgroups}

We now see an application of our classifier in an applied problem faced by Bank
of Italy. We compare the best performing classifier (\COMB) with
a method commonly used to estimate the probability of
one-year default by companies at aggregate level.

Consider a segmentation of all the companies (e.g., according to
economic sector, geographical area, etc.). Often there is the need to
estimate the \emph{probability of default} (PD) of a loan in a given segment.
A very simple approach, which is actually used in practice, is
to simply take the ratio of the companies in the segment that went into
default at year $T+1$ over all the companies that were not in default in
year $T$. We use this method as a baseline.

We now consider a second approach based on our classifier, which we call
\COMB. We estimate the PD by considering the amount of companies in the
segment that are
expected (using the \COMB classifier) to go into
default at year $T+1$  compared to the total loans
existing for the segment at the time $T$.

We use two different segmentations. A coarse one, in which the segments
are defined by the economic sector (e.g. mineral extraction,
manufacturing) .
and a finer one, which is defined by the combination of the economic
sector and the geographic area, as defined by a value similar to the
company's zip code.

In Table~\ref{tbl:comp_pd} we compare the two approaches for estimating
the~PD.
As expected, in both segmentations the classifier-based approach is a
winner, with the improvement being larger for the finer segmentation.
In many cases the two approaches give the same result, typically because
in these cases there are no companies that fail (PD equals 0).

\setlength{\tabcolsep}{6pt}
\begin{table}[]
\small
\flushleft
\begin{tabular}{llllllll}
\cline{1-3} \cline{5-8}
\multicolumn{3}{l}{\textbf{Coarse segmentation}} &  & \multicolumn{3}{l}{\textbf{Fine segmentation}} & \\
&\textbf{Baseline}         &\COMB        &  & &\textbf{Baseline}         &\COMB        &  \\ \cline{1-3} \cline{5-8} \\
\multicolumn1{p{25mm}}{\textbf{Mean error}}           & 0.11              & 0.048            &  &\multicolumn1{p{25mm}}{\textbf{Mean error}}                 & 0.088              & 0.036            &  \\
\multicolumn1{p{25mm}}{\textbf{Var error}}                  & 0.056             & 0.016            &  &\multicolumn1{p{25mm}}{\textbf{Var}}                & 0.06             & 0.025            &  \\
\multicolumn1{p{25mm}}{\textbf{Superiority percentage} }   & 25.1\% &
45.6\%      &  &\multicolumn1{p{25mm}}{\textbf{Superiority percentage} } & 6.1\%             & 19.5\% \\
\hline
\end{tabular}\\
\caption{Comparison of the standard approach to estimate PD with the
classifier-based one. ``Mean error'' is the average error between the
predicted PD value and the real one. ``Var error'' is the variance of the
error. ``Superiority percentage'' is the percentage of segments in
which the predictor is better than the other; in the remaining ones we have the
same performance.}
\label{tbl:comp_pd}
\end{table}
%
%

\section{Conclusion}
\label{sec:conclusion}

Business-failure prediction is a very important topic of study for
economic analysis and the regular functioning of the financial system.
Moreover the importance of this issue has greatly increased following
the recent financial crisis. There have been many recent studies that
have tried to predict the failure of companies using various
machine-learning techniques.

In our study, we used for the first time credit information from the
Italian Central Credit Register to predict the banking default
of Italian companies, using Machine Learning techniques. We analyzed a
very large dataset containing information about \emph{almost all the loans of
all the Italian companies}. Our first findings is that, as in the case
of bankruptcy prediction, machine-learning approaches are able to
outperform significantly simpler statistical approaches.
Moreover, combining classifiers of different type can lead to even better results.
Finally, using
information on past loan data is crucial, but the additional use of
balance-sheet data can improve classification even further.

We show that the combined use of loan data with balanced-sheet data
leads to improved performance for predicting default. We conjecture that
using loan data in the prediction of bankruptcy (where, typically, only
balance-sheet data are being used) can improve further the performance.

Nevertheless, prediction remains an extremely hard problem. Yet,
even slight improvement in the performance, can lead to savings of
multiple hundreds of euros for the banking system. Thus our goal is to
improve classification even further by combining our approaches with
further techniques, such as neural-network based ones. Some preliminary
results in which we use only neural networks are encouraging, even
though are worse than the results we report here.

\bibliographystyle{splncs04}
\bibliography{finance}

\end{document}